\documentclass[journal]{IEEEtran}

\usepackage{graphicx}
\usepackage{lineno}
\usepackage{stfloats}
\usepackage{xcolor}
\usepackage{ragged2e}
\usepackage{amsmath}
\usepackage{tabularray}
\usepackage{longtable}
\usepackage{supertabular}
\usepackage{tabularx}
\usepackage{booktabs} 
\usepackage{xcolor}
\usepackage{textgreek}

\ifCLASSINFOpdf

\else

\fi

\usepackage{graphicx}
\usepackage{lineno}
\usepackage{stfloats}
\usepackage{multicol}
\usepackage{array}
\usepackage{verbatim}
\usepackage{amsmath}
\usepackage{tabularray}
\usepackage{longtable}
\usepackage{supertabular}
\usepackage{tabularx}
\usepackage{booktabs} 
\usepackage{xcolor}
\begin{document}

	\title{White Gaussian Noise Generation with a Vacuum State Quantum Entropy Source Chip}

	\author{Guan-Ru~Qiao,
			Bing~Bai,
			Zi-Xuan~Weng,
			Jia-Ying~Wu,
			You-Qi~Nie,
			and Jun~Zhang

			\thanks{Guan-Ru Qiao, Bing Bai, Zi-Xuan Weng, Jia-Ying Wu are with Hefei National Research Center for Physical Sciences at the Microscale and School of Physical Sciences, University of Science and Technology of China, Hefei 230026, China, and are also with CAS Center for Excellence in Quantum Information and Quantum Physics, University of Science and Technology of China, Hefei 230026, China. Corresponding author: Bing Bai (e-mail: baib@ustc.edu.cn).}
			\thanks{You-Qi Nie and Jun Zhang are with Hefei National Research Center for Physical Sciences at the Microscale and School of Physical Sciences, University of Science and Technology of China, Hefei 230026, China, and with CAS Center for Excellence in Quantum Information and Quantum Physics, University of Science and Technology of China, Hefei 230026, China, and also with Hefei National Laboratory, University of Science and Technology of China, Hefei 230088, China. Corresponding author: Jun Zhang (e-mail: zhangjun@ustc.edu.cn).}}
			\maketitle

	\begin{abstract}
		White Gaussian noise (WGN) is widely used in communication system testing, physical modeling, Monte Carlo simulations, and electronic countermeasures.
		WGN generation relies heavily on random numbers.	
		In this work, we present an implementation of WGN generation utilizing a quantum entropy source chip for the first time.
		A photonic integrated chip based on the vacuum state scheme generates quantum random numbers at a real-time output rate of up to 6.4 Gbps.
		A hardware-based inversion method converts uniform quantum random numbers into Gaussian random numbers using the inverse cumulative distribution function.
		Subsequently, the WGN signal is generated through a digital-to-analog converter and amplifiers.
		The WGN generator is characterized by a bandwidth of 230 MHz, a crest factor as high as 6.2, and an adjustable peak-to-peak range of 2.5 V.
		This work introduces a novel approach to WGN generation with information-theory provable quantum random numbers to enhance system security.
		\end{abstract}


	\section{Introduction}
		\IEEEPARstart
		White Gaussian noise (WGN) is a standard model representing the cumulative effects of various random noise sources.
		A white Gaussian noise generator (WGNG) is widely used to produce WGN, which plays a crucial role in physical modeling, Monte Carlo simulations \cite{bucklew2003monte}, and the evaluation of communication systems \cite{wyner1967probability, lee2003hardware}, particularly for testing bit error rates \cite{burr1997block, binti2006effect} and signal-to-noise ratios.
		WGN is also extensively applied in electronic countermeasures, including electronic interference \cite{le2007noise} and radar countermeasures \cite{mooney1998robust, paik2014effectiveness}.
				
		WGNGs can be implemented using physical noise sources, such as the thermal noise or breakdown noise of a diode, whose noise signal is not strictly Gaussian-distributed. 
		Subsequent filtering and amplification are typically required to approximate WGN.
		Digital hardware implementations of WGNGs provide well-distributed properties along with adjustable noise bandwidth and amplitude.
		Such WGNGs integrate a Gaussian random number (GRN) generator, which relies on uniform random numbers (URNs) as a crucial resource for generating WGN \cite{thomas2007gaussian, malik2016gaussian}.
		The quality of random numbers directly impacts the randomness, distribution, and bandwidth of the noise.
		Extensive research has focused on developing hardware-based approaches for generating GRNs, including the Box-Muller method \cite{lee2003hardware, box1958note, alimohammad2005area}, rejection-acceptance methods represented by the Ziggurat technique \cite{marsaglia2000ziggurat, zhang2005ziggurat}, the central limit theorem method \cite{thomas2014fpga, chen2024flexible}, and the inversion method \cite{lee2006inversion, cheung2007hardware, echeverria2007fpga, de2010new, gutierrez2012hardware, choi2016area}.
		
		Classical methods of generating WGN typically rely on pseudo-random numbers.
		However, pseudo-random number generators have inherent limitations, such as predictability, periodicity, and seed security, which present significant risks.
		These limitations undermine the accuracy of simulations and tests, particularly in high-security and unpredictable applications, such as electronic countermeasures.
		Predictable noise patterns in these scenarios may introduce system vulnerabilities, reducing the effectiveness of noise-based interference.
		To address these problems, quantum physics offers an effective solution.
		Quantum random number generators (QRNGs) produce true random numbers with characteristics of unpredictability, irreproducibility, and unbiasedness, which are guaranteed by the fundamental principles of quantum mechanics.
		WGNGs incorporating quantum random numbers provide a secure solution for generating WGN in hardware systems.

		Over the past two decades, various QRNG schemes have been proposed and demonstrated \cite{herrero2017quantum, ma2016quantum} including the beam splitter scheme by measuring the path selection of single photons \cite{stefanov2000optical, jennewein2000fast}, the time measurement scheme by digitizing the arrival time of single photons \cite{dynes2008high, nie2014practical}, the quantum phase fluctuation scheme by measuring phase fluctuations due to the spontaneous emission of laser \cite{xu2012ultrafast, nie2015generation, zhang2016note}, and the vacuum state scheme by measuring quantum noise fluctuations \cite{gabriel2010generator, bai202118, bruynsteen2023100, tanizawa2024real}.
		Among these methods, the vacuum state scheme stands out for its high-speed advantages and relatively fewer component requirements, making it particularly suitable for integration \cite{bai202118}.
		
		\begin{figure*}[hb]
			\centerline{\includegraphics[width=18 cm]{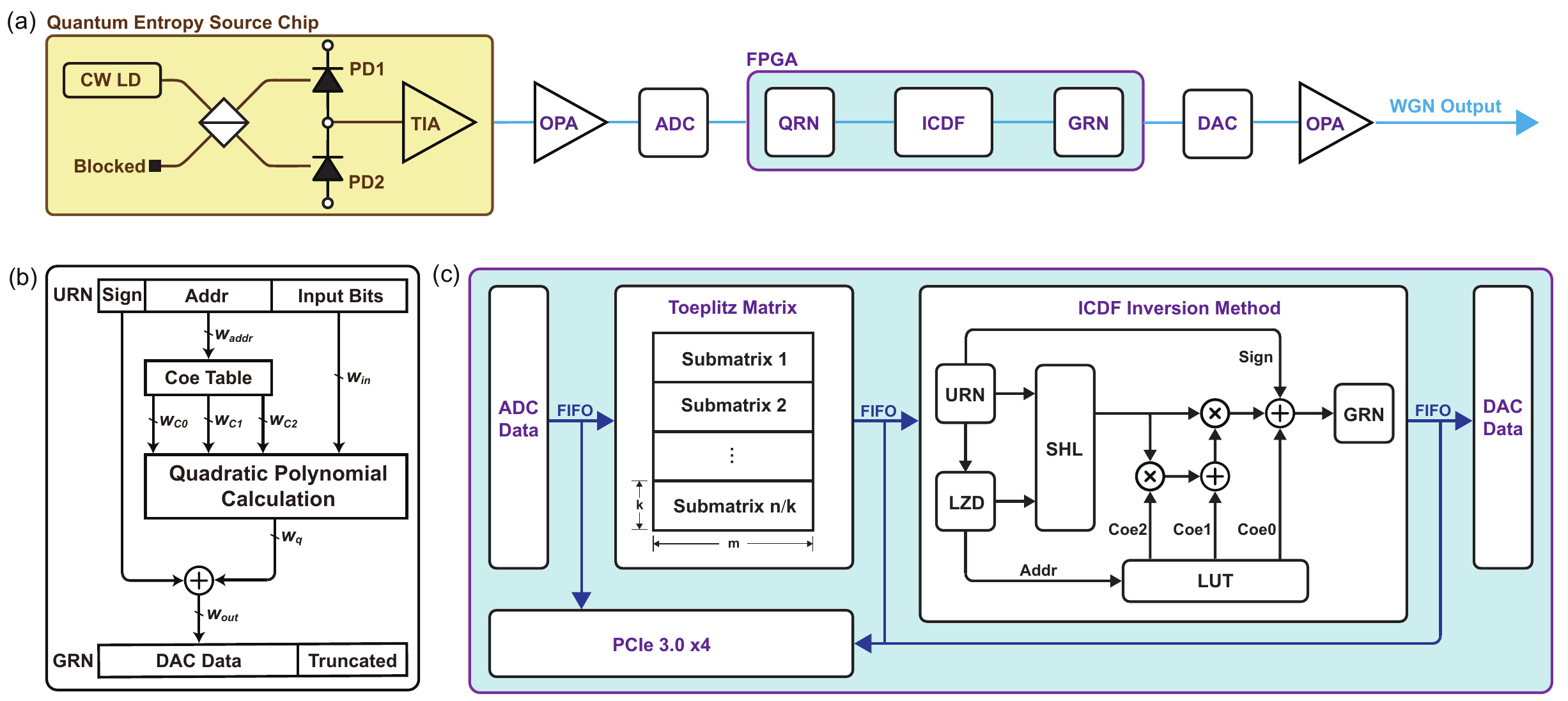}}
			\caption{(a) The schematic of the WGNG system. (b) The architecture of the ICDF inversion method algorithm. (c) The WGNG data processing diagram. WGNG: white Gaussian noise generator, CW LD: continuous wave laser diode, PD: photodetector, TIA: transimpedance amplifier, OPA: operational amplifier, ADC: analog-to-digital converter, QRN: quantum random number, ICDF: inverse cumulative distribution function, GRN: Gaussian random number, DAC: digital-to-analog converter, Addr: address, Coe: coefficient URN: uniform random number, LZD: leading zero detector, SHL: left barrel shifter, FIFO: first input first output.}
			\label{figure1}
			\end{figure*}

		In this work, we present a WGNG based on a vacuum state quantum entropy source chip for the first time.
		Quantum random numbers are generated at a real-time rate of 6.4 Gbps using a vacuum state photonic integrated chip.
		The quantum URNs are transformed into GRNs using the inverse cumulative distribution function (ICDF) of a Gaussian distribution through an inversion method implemented in a field-programmable gate array (FPGA).
		The WGN signal is generated from GRNs via a digital-to-analog converter (DAC) and then output after passing through operational amplifiers (OPAs).
		The entire WGNG system is compactly housed in a 2U case, with dimensions of 37 $\times$ 31 $\times$ 11 cm³.
		After characterization, the WGNG achieves a bandwidth of 230 MHz and a crest factor (CF) of 6.2, with an adjustable peak-to-peak range of 2.5 V.

	\section{Implementation of the WGNG}

		Fig.~\ref{figure1}(a) illustrates the schematic of the WGNG system, which transforms quantum-generated randomness into an amplitude-adjustable white Gaussian noise output.
		The vacuum state quantum entropy source chip serves as the core component of the QRNG system.
		The quantum noise signal is amplified in two stages by OPAs to ensure that the signal meets the measurement requirements of the analog-to-digital converter (ADC).
		The digital data from the ADC is then transmitted to the FPGA for further processing.		
		This step converts the raw quantum-generated data into high-quality random numbers via randomness extraction, preparing them for conversion to a Gaussian distribution.
		The processed random numbers are transferred to the GRN calculation unit of the FPGA.
		The inversion method is applied to transform the quantum URNs into Gaussian-distributed values via the Gaussian ICDF.		
		Finally, the GRNs are sent to the DAC, producing WGN.
		The resulting WGN signal is output with adjustable amplitude through OPAs.

	\subsection{Quantum Entropy Source Chip}
		In the vacuum state QRNG scheme, the quantum noise fluctuations are measured using homodyne detection.
		The randomness arises from the quadrature measurement on the vacuum state, with the uncertainty relation guaranteeing the unpredictability of the measurement outcomes.
		The local oscillator (LO) light is injected into the input port of a 50:50 beam splitter with the other port blocked.
		In homodyne detection, the direct current is canceled out, and Gaussian-distributed vacuum fluctuations are obtained.

		The vacuum state method employs photodetectors (PDs) instead of highly sensitive single-photon detectors, allowing the system to achieve much higher random bit rates, which makes it especially suitable for rapidly generating large volumes of random data.
		Another key advantage of the vacuum state method is its compatibility with photonic integration technology.
		As a quantum entropy source, vacuum state noise fluctuations eliminate the need for bulky external components, enhancing the integration of the QRNG system.
		The system’s optical components can be fully integrated onto a single photonic chip, substantially reducing its structural complexity.
		These features make it a highly effective solution for QRNG-based Gaussian noise generation.
		
		We designed a photonic integrated quantum entropy source chip based on vacuum state fluctuations and implemented it in a butterfly package \cite{bai202118}.
		The photonic integration offers a compact and efficient solution while ensuring high-quality random number generation performance.
		A continuous-wave laser diode, acting as the LO, interferes with the vacuum state at a 50:50 beamsplitter in the waveguide, resulting in two output beams with balanced power levels.
		The laser diode operates at a center wavelength of 1550 nm.
		The homodyne detection signal is extracted through the common electrode of the two PDs.
		The PDs exhibit a responsivity of 0.9 A/W at a wavelength of 1550 nm.
		The transimpedance amplifier (TIA) is integrated into the photonic chip via wire bonding to minimize parasitic parameters and amplifies the detected photocurrent difference from the PD module.			
		For system stability and reliability, control of the laser diode and TIA, along with current monitoring of the PDs, is managed via the FPGA and dedicated monitoring chips.
		This configuration ensures the consistent generation of quantum noise signals, providing a reliable quantum entropy source for the WGNG system.

	\subsection{Inversion Method}
		A transformation algorithm is implemented in the FPGA to convert URNs into GRNs effectively.
		However, many algorithms face challenges when generating GRNs based on a quantum entropy source.
		The Box-Muller method efficiently transforms pairs of URNs into GRNs, providing consistent output rates and relatively high throughput.
		However, it demands significant hardware resources and extensive optimization to compute trigonometric and logarithmic functions efficiently.
		Among rejection-acceptance methods, the Ziggurat technique is particularly notable for its application in software-based random number generation.
		The Ziggurat technique encounters difficulties in maintaining a constant GRN output due to the rejection of certain samples.
		Even with efficient hardware solutions like GRN buffering, inherent variability can restrict throughput in large-scale data streams.
		The central limit theorem states that the sum of a sufficiently large number of independent, uniformly distributed random variables approximates a normal distribution.
		The central limit theorem method approximates a Gaussian distribution using a weighted sum of smaller component distributions combined with alias table read-only memories.
		This requires numerous URN samples with high data bit widths to achieve high-accuracy GRNs, which makes the approach impractical for systems constrained by a quantum entropy source with limited URNs.

		Compared to the methods mentioned above, the inversion method has unique advantages.
		It is a widely used hardware approach that transforms URNs into GRNs through the ICDF. 
		The inversion method stands out as a general technique for generating various probability distributions by leveraging the ICDF with uniform random variables, including the standard Gaussian distribution, as described in~\eqref{eq1}.
		The ICDF of the standard Gaussian distribution can be represented as in~\eqref{eq2}, where $\mathrm{erf}^{-1}(x)$ denotes the inverse error function.
			\begin{align}
				\Phi(x) = \frac{1}{\sqrt{2\pi}} \int_{-\infty}^x e^{-\frac{x^2}{2}} \, dx \label{eq1} \\
				\Phi^{-1}(x) = \sqrt{2} \, \mathrm{erf}^{-1}(2x - 1) \label{eq2}
			\end{align}

		In FPGA-based implementations, the core task is efficiently computing the ICDF, where the inherent parallelism of FPGAs and their fixed-point arithmetic offer significant advantages.
		The inversion method utilizes adders, multipliers, and lookup tables (LUTs) that store polynomial coefficients in the FPGA.
		To achieve higher bandwidth for WGN, the inversion method generates a rapid and stable GRN data stream that matches the dedicated bit width of the DAC. 
		This optimizes the utilization of the constrained quantum random number resources.
		LUTs and piecewise fitting techniques are employed to meet computational demands.

		Specifically, a non-uniform segmentation scheme combined with polynomial fitting approximates the ICDF function, as depicted in Fig.~\ref{figure1}(b).
		The URN data is divided into three components: the sign bit, the address bits, and the input bits for computation.
		The segmentation of the ICDF depends on the width of the URN output data.
		A larger URN bit width improves the ICDF computational accuracy but requires more polynomial fitting segments.
		Since the ICDF of a Gaussian distribution is an odd function centered at 0.5, only half the fitting segments are required.
		As described in \cite{cheung2007hardware}, segmenting the ICDF into a geometric sequence based on powers of 2 is sufficient for precise curve fitting.
		Each segment is further subdivided into four parts to enhance accuracy.
		The polynomial coefficients derived from this segmentation are adjusted to the appropriate bit width and preloaded into the LUT.
		Input bits, determined by the coefficient table address, are then sent to the quadratic polynomial calculation module.
		By incorporating the sign bit into the final result, this process effectively transforms uniformly distributed quantum random numbers into Gaussian random numbers with high precision.

	\subsection{Data Processing}
		Fig.~\ref{figure1}(c) illustrates the data flow in the FPGA, showcasing the transformation of quantum randomness into usable WGN output.
		The quantum noise signal captured by the ADC is fed into the FPGA.
		The FPGA plays a crucial role in processing this randomness and converting the continuous stream of signals into a series of URNs through real-time randomness extraction algorithms.
		The Toeplitz matrix serves as a hashing extractor to refine the raw randomness \cite{zhang2016note, ma2013postprocessing}.
		For a Toeplitz matrix with dimensions $m\times n$, where $m$ represents the number of final extracted random bits and $n$ represents the number of raw bits, the FPGA multiplies the matrix by $n$ raw bits to extract $m$ bits of randomness.
		Given the computational constraints of the FPGA, direct processing of large matrices is infeasible.
		To address this, the Toeplitz matrix is divided into smaller submatrices.
		The matrix dimensions are partitioned into $n/k$ submatrices, enabling the FPGA to process smaller data blocks in parallel, enhancing overall system efficiency.
		The raw data and quantum random numbers can be transmitted to a host computer.
		The PCIe 3.0 x4 interface ensures the system meets data throughput requirements without introducing bottlenecks.

		Before the generation of GRN data, the quantum random numbers are processed through a first in first out (FIFO) buffer to meet a dedicated URN bit width requirement. 
		The input and output clocks of the FIFO are driven by a phase-locked loop, which aligns the clock frequency and URN data bit width with the data rate, ensuring complete utilization of all quantum random numbers.
		Implementing the GRN generator within the FPGA involves several critical components: a leading zero detector, a left barrel shifter for logical shifts, adders, multipliers, and LUTs storing polynomial fitting coefficients.
		The process begins with the leading zero detector, which identifies the position of the most significant zero bit in the input URN data.
		This position determines the segment number corresponding to the count of leading zeros.
		Additionally, the two bits immediately following the most significant one bit are extracted to select the address of one of the four segment partitions.
		The segment number and URN data are then passed to the left barrel shifter module, where the data undergoes a left-shift operation.
		This step reduces the bit width of the data, optimizing computational efficiency.
		The truncated URN value is forwarded to the polynomial fitting module.
		A quadratic polynomial piecewise fitting approach is employed, which has been shown to meet the precision requirements of the ICDF \cite{lee2006inversion, gutierrez2012hardware}.
		This method balances computational efficiency and precision, ensuring the statistical validity of the output random numbers.

		To achieve the desired precision, the bit width of the truncated input random number $w_{in}$ is set to 11 bits.
		The polynomial coefficients are assigned the following bit widths: $w_{c0}$ = 35 bits, $w_{c1}$ = 20 bits, and $w_{c2}$ = 7 bits.
		The output of the polynomial fitting module has a bit width $w_{out}$ of 35 bits, which is subsequently truncated to 13 bits.
		Finally, after incorporating the sign bit from the original data, the system outputs a 14-bit GRN, which is transmitted to the DAC.	

	\subsection{WGN Output}

		\begin{figure*}[ht]
			\centerline{\includegraphics[width=17 cm]{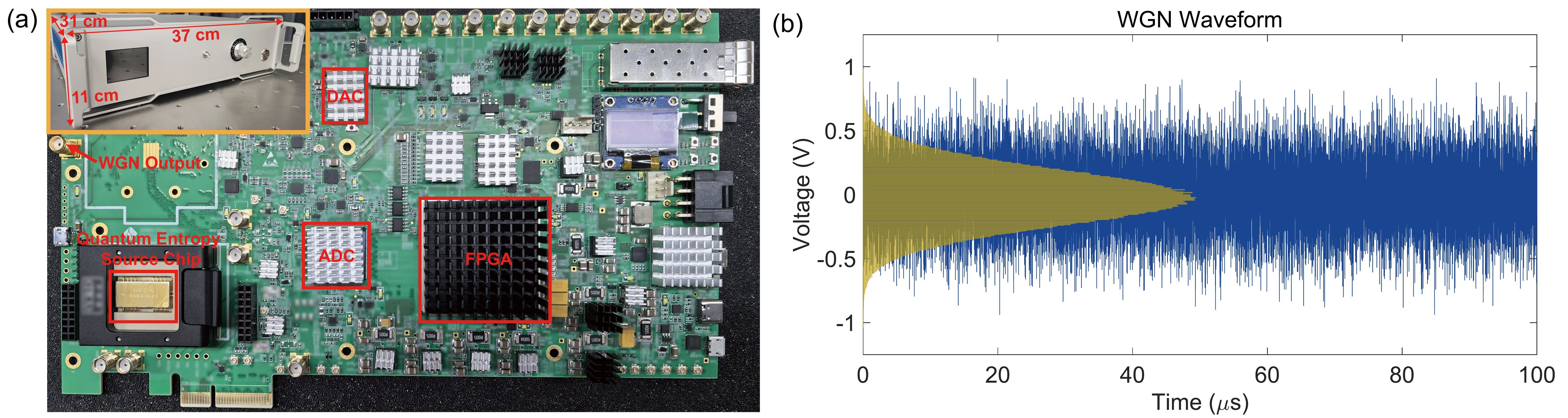}}
			\caption{(a) The photo of the WGNG system. (b) The typical WGN waveform with a duration of 100 µs and the statistical distribution of the WGN with 12-bit URNs input.}
			\label{figure2}
			\end{figure*}

		The bandwidth and CF of the WGN are configurable across four scales, corresponding to different URN input bit widths: 12, 16, 24, and 32 bits.
		The WGN is generated by outputting GRN data through a 14-bit DAC.
		Using a balun, the differential WGN signal from the DAC output is converted into a single-ended signal.
		After passing through two adjustable amplifiers, the final WGN signal is obtained.

		The photo of the WGNG system is shown in Fig.~\ref{figure2}(a).
		The entire WGNG system is integrated onto a printed circuit board and compactly housed in a 2U case measuring 37 $\times$ 31 $\times$ 11 cm³.
		The WGN signal is output through a subminiature port.
		Communication between the host computer and the system is facilitated via a serial port or a PCIe interface, allowing state monitoring and command issuance.
		The amplitude of the WGN signal can be adjusted using a knob or, alternatively, via the host computer.

	\section{Characterization of WGNG}

		\subsection{Quantum Randomness}
			The quantum noise generated by the quantum entropy source chip follows a Gaussian distribution \cite{bai202118}.
			To optimize the performance of the system, the TIA is configured with an output bandwidth spanning 1 MHz to 1.5 GHz and a transimpedance gain of 5 k$\Omega$.
			Considering the frequency response characteristics of the quantum entropy source chip, an 8-bit ADC sampling rate of 1.2 GSa/s is employed to capture high-speed quantum noise data accurately.
			To evaluate the randomness in the experiment, given that both quantum and classical noise follow Gaussian distributions, the total variance of the measured signal amplitude is expressed as~\eqref{eq3}.
				\begin{equation}\label{eq3}
					\sigma_{total }^{2}=\sigma_{q}^{2}+\sigma_{c}^{2}
				\end{equation}
			Here, $\sigma_{total }^{2}$ is the total variance of the amplitude measured by the ADC, $\sigma_{q }^{2}$ is the quantum noise contribution, and $\sigma_{c }^{2}$ is the classical noise contribution.
			$\sigma_{c }^{2}$ is determined under conditions without the LO input.
			When the power of the LO is set to 7.05 mW, the chip operates optimally, producing the original quantum noise signal with $\sigma_{q }^{2}$ = 767.4 and $\sigma_{c }^{2}$ = 7.9.
			The randomness of the raw data is evaluated by a min-entropy approach, calculated as shown in~\eqref{eq4}.
				\begin{equation}\label{eq4}
					H_{min }(X)=-\log _{2} P_{max }
				\end{equation}
			The min-entropy, $H_{min }(X)$, is 5.97 bits per sample.
			The extracted randomness comprises both the intrinsic randomness originated from quantum shot noise and the nominal randomness from mixed states, where the latter might be controlled by classical or quantum side information \cite{zhou2018randomness, yuan2019quantum}.
			In principle, one can take account of the potential leakage of side information and quantify the amount of intrinsic randomness with condition min-entropy.

			\begin{table}[h!]
				\centering
				\caption{The typical NIST test result for 1 Gb of quantum random data}
				\label{t1}
				\begin{tabular}{lcc}
					\toprule
					Statistical Test          & Proportion & P-Value \\ 
					\midrule
					Frequency                 & 0.992      & 0.352   \\
					Block Frequency           & 0.988      & 0.307   \\
					Cumulative Sums           & 0.993      & 0.333   \\
					Runs                      & 0.993      & 0.910   \\
					Longest Run               & 0.986      & 0.315   \\
					Rank                      & 0.990      & 0.481   \\
					FFT                       & 0.986      & 0.522   \\
					Non-overlapping Template  & 0.991      & 0.915   \\
					Overlapping Template      & 0.989      & 0.685   \\
					Universal                 & 0.991      & 0.389   \\
					Approximate Entropy       & 0.991      & 0.386   \\
					Random Excursions         & 0.989      & 0.587   \\
					Random Excursions Variant & 0.989      & 0.906   \\
					Serial                    & 0.993      & 0.467   \\
					Linear Complexity         & 0.992      & 0.990   \\
					\bottomrule 
					
				\end{tabular}
			\end{table}
			
			The raw data, at a rate of 9.6 Gbps, is fed into the FPGA in parallel for further processing.
			Given the matrix dimensions, the extraction ratio for the Toeplitz matrix is 3:2 based on the chip's $H_{min }(X)$, with submatrix parameters set to $m$ = 1024, $n$ = 1536, and $k$ = 64.
			The 9.6 Gbps raw data and 6.4 Gbps quantum random numbers can be transmitted to the host computer via a PCIe 3.0 x4 interface, supporting a transfer rate of 10 Gbps. 
			A typical NIST test \cite{NIST} result for a quantum random data file of size 1 Gb is presented in Table~\ref{t1}.
			All p-values exceed 0.01, and all proportions exceed 0.98, indicating that the random bits successfully pass the NIST tests.
			The final random numbers pass all test items, confirming the statistical validity of the output randomness and its suitability as a seed for GRN generation.

		\subsection{Distribution of the WGN}
		
			\begin{figure*}[t!]
				\centerline{\includegraphics[width=17 cm]{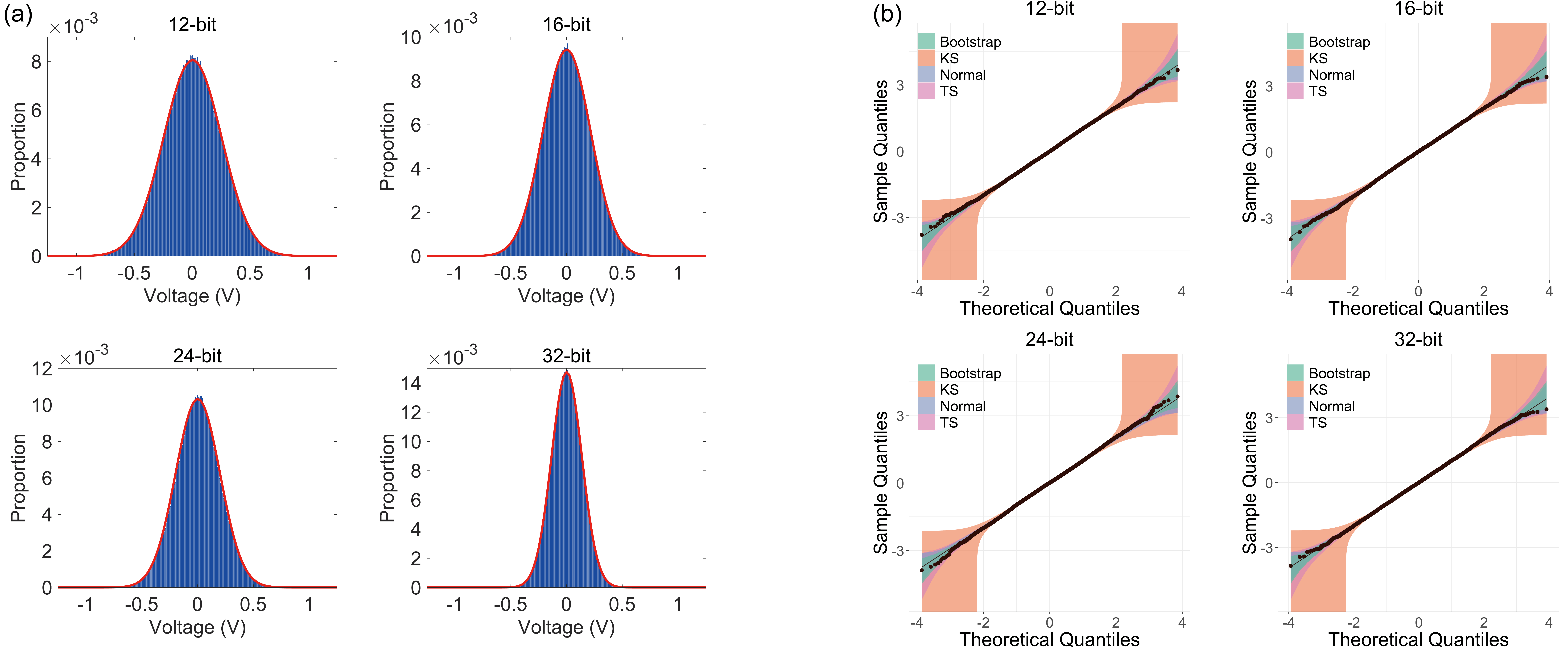}}
				\caption{(a) The histograms and theoretical Gaussian distribution fit of WGN for all four configurations. Each file consists of $10^6$ samples. (b) The quantile-quantile plots of WGN for all four configurations. Each file consists of $10^4$ samples. KS: Kolmogorov-Smirnov method, TS: tail-sensitive method.}
				\label{figure3}
				\end{figure*}

			The amplitude of noise signals following a Gaussian distribution is a critical characteristic of WGN.
			The waveform diagram of typical WGN is shown in Fig.~\ref{figure2}(b).
			The maximum output range of the generated GRN data is determined by the maximum value of the ICDF for a given input bit width.
			The QRNG data is configured for four different bit widths: 12, 16, 24, and 32 bits.
			Statistical WGN histograms for the four data bit widths are shown in Fig.~\ref{figure3}(a).
			These data were collected using an oscilloscope with $10^6$ samples.

			The distribution of WGN signal data can be evaluated using various detection tools beyond observing its distribution shape and fitting curve.
			A powerful statistical method for evaluating the quality of generated noise is the quantile-quantile (Q-Q) plot \cite{pleil2016qq}.
			This plot compares the quantiles of WGN data to those of a Gaussian distribution, allowing a visual evaluation of their similarity to the target Gaussian distribution.
			The Q-Q plot serves as a critical tool for detecting deviations from normality in the tails of Gaussian distributions, whereas histograms are more effective for analyzing the central regions of a distribution.
			In a Q-Q plot, the horizontal and vertical axes represent the quantiles of the two distributions being compared.
			When two distributions are identical, their points on the Q-Q plot align closely with a straight line, known as the reference line.
			Confidence interval lines on the Q-Q plot indicate a range of values where a specified percentage of data points is expected to fall.
			
			Fig.~\ref{figure3}(b) shows Q-Q plots for QRNG data at different bit widths, each based on a statistical sample size of $10^4$.
			These plots include four distinct confidence intervals derived from different statistical approaches: normal pointwise confidence bands, the bootstrap method \cite{horowitz2018bootstrap}, the Kolmogorov-Smirnov method, and the tail-sensitive method \cite{aldor2013power}.
			Additionally, the WGN data has passed the Lilliefors test and Jarque-Bera test, which are standard Gaussian distribution tests for large sample sizes \cite{razali2010power}.
			All confidence intervals in these tests are set to 95\%, a commonly adopted standard in statistics that balances result precision with the confidence needed to support decisions or inferences.

		\subsection{Bandwidth and Crest Factor}

\begin{figure*}[htp]
				\centerline{\includegraphics[width=17 cm]{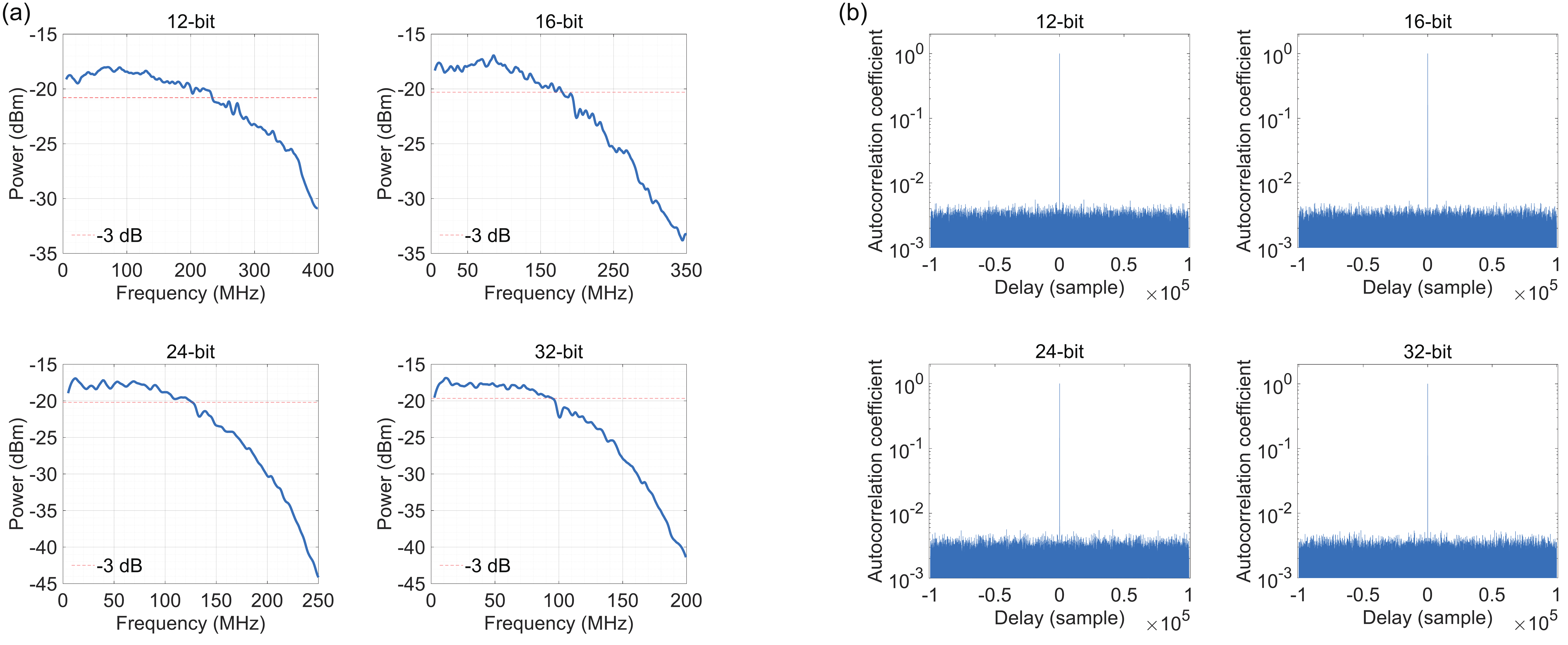}}
				\caption{(a) The power spectral density curve of WGN for all four configurations. (b) The absolute value of the autocorrelation coefficient of WGN for all four configurations. Each file consists of $10^6$ samples.}
				\label{figure4}
				\end{figure*}

			Bandwidth and noise flatness are critical parameters for evaluating WGN.
			Fig.~\ref{figure4}(a) shows the smoothed power spectral curves of the WGN with different bit widths, measured using a spectrum analyzer.
			Noise flatness is defined as the uniformity of the power spectral density of noise across the bandwidth. 
			It indicates whether the noise power is evenly distributed across different frequencies.
			Ideally, perfect WGN exhibits an infinitely wide, absolutely flat power spectrum, as theoretically predicted.
			However, practical limitations, such as the data rate of the URN source, the operating frequency of the DAC, and the properties of the amplification circuit, limit the bandwidth range of a WGN to a finite spectrum.
			When the QRNG produces data at a certain rate, selecting a higher bit-width URN requires reducing the GRN computation clock frequency, which in turn directly impacts the signal bandwidth of the generated WGN.

			The CF is another key WGN parameter that reflects the range of the distribution tails, which is crucial in signal interference testing.		
			The CF is defined as the ratio of the peak value to the root mean square (RMS) value of the noise signal, as shown in~\eqref{eq5}, where $V_{pp }$ denotes the peak-to-peak value \cite{xu2012parallel}.
			\begin{equation}\label{eq5}
					C F=\frac{V_{Peak }}{V_{RMS }}=\frac{\frac{1}{2} V_{pp }}{\sigma}
				\end{equation}
			In WGN, the RMS equals the standard deviation $\sigma$, as the mean value $\mu$ = 0.
			The inversion method is based on the standard Gaussian ICDF.
			The bit width of URNs determines the mapping range of GRNs.
			Consequently, the maximum value of normalized GRNs is the value of the CF, calculated using~\eqref{eq6}, where $w$ refers to the input URN bit width.
				\begin{equation}\label{eq6}
					C F= \sqrt{2} \, \mathrm{erf}^{-1}(1 - \frac{1}{2^{w-1} } )
				\end{equation}
			Increasing the input URN bit width reduces the clock frequency of the GRN computation module.
			The bandwidth of the WGN and its corresponding CF values show a negative correlation, as presented in Table~\ref{t2}.
			
			\begin{table}[htp]
				\centering
				\caption{The bandwidth and CF of the WGN}
				\label{t2}
				\begin{tabular}{lcccc}
				\toprule
				URN Bit Width   & 12-bit & 16-bit & 24-bit & 32-bit  \\
				\midrule
				Bandwidth (MHz) & 230    & 175    & 125    & 85      \\
				Calculated CF   & 3.5    & 4.2    & 5.3    & 6.2   	 \\
				\bottomrule
				\end{tabular}
			\end{table}

		\subsection{Autocorrelation Coefficient}
			Statistical independence is an essential characteristic of WGN.
			An autocorrelation coefficient test is conducted to evaluate the intrinsic randomness of the resulting WGN data, as shown in Fig.~\ref{figure4}(b).
			The autocorrelation function measures the dependence of a signal at different time points.
			Practical WGN generated within a limited bandwidth may exhibit slight autocorrelation due to bandwidth limitations.
			The absolute autocorrelation coefficients for all samples are below 0.01, except for the region near zero delay, which indicates nearly non-existent autocorrelation.
			The autocorrelation curves for all four input bit-width configurations show no discernible patterns, confirming the statistical independence of the noise samples.

	\section{Conclusion}
		In summary, we have reported, for the first time, a WGNG equipped with a vacuum state quantum entropy source chip.
		The photonic integrated quantum entropy source chip achieves a real-time quantum random number bit rate of 6.4 Gbps.
		An inversion method based on the ICDF of a Gaussian distribution is implemented in an FPGA, efficiently converting quantum URNs into GRNs, and sent to a DAC for WGN generation.
		The WGNG exhibits an adjustable peak-to-peak output range of 2.5 V, achieving a bandwidth of 230 MHz and a CF of 6.2, with dimensions of 37 $\times$ 31 $\times$ 11 cm³.
		This work presents a novel approach to WGN generation using quantum randomness that is highly suitable for applications requiring high security, such as electronic countermeasures.


\ifCLASSOPTIONcaptionsoff
  \newpage
\fi


\bibliography{wgn}
\bibliographystyle{IEEEtran}

\end{document}